\documentclass[12pt]{article}
\usepackage{amsmath}
\usepackage{amssymb}
\usepackage{amsfonts}
\usepackage[T2A]{fontenc}
\usepackage[cp1252]{inputenc}
\usepackage[english]{babel}
\usepackage[centerlast,small]{caption2}
\usepackage{graphicx}

\newcommand{\intk}{\displaystyle \int\frac{d^D k}{(2\pi)^D}}

\newcommand{\la}{\lambda}
\newcommand{\lnq}{\log\left\vert q^2\right\vert}
\newcommand{\gamm}{\tilde{\gamma}}

\begin{document}

\begin{titlepage}

\begin{center}

{\bf \large Quantum corrections to spin effects in general relativity}

\vspace{1cm}

G.G. Kirilin\footnote{g\_kirilin@mail.ru}

\vspace{1cm}

Budker Institute of Nuclear Physics\\
630090 Novosibirsk, Russia

\end{center}

\bigskip

\begin{abstract}
Quantum power corrections to the gravitational spin-orbit and spin-spin
interactions, as well as to the Lense-Thirring effect, were found for particles
of spin $1/2$. These corrections arise from diagrams of second order in Newton
gravitational constant $G$ with two massless particles in the unitary cut in
the $t$-channel. The corrections obtained differ from the previous calculation
of the corrections to spin effects for rotating compound bodies with spinless
constituents.

\bigskip


\end{abstract}

\vspace{8cm}

\end{titlepage}

\section{Introduction}
It is common knowledge that the scattering amplitude
$\mathcal{M}=\langle1(p_1-q),2(p_2+q)\vert1(p_1),2(p_2)\rangle$ of two
gravitating bodies contains terms logarithmic in the momentum transfer $\lnq$
in the $q^2\to 0$ limit. The terms arise from diagrams with two massless
particles in the unitary cut in the t-channel. In the nonrelativistic limit
these terms generate quantum power corrections to the classical Hamiltonian
\cite{bode}. Thus quantum corrections to the Newton potential take the
form~[1-20]

\begin{align}
U = -\,\frac{41}{10\pi}\,\frac{G^2\hbar\,m_1 m_2}{c^3
r^3}-\,\frac{4}{15\pi}\,\frac{G^2\hbar\,m_1 m_2}{c^3
r^3}-\,\frac{N_\nu}{15\pi}\,\frac{G^2\hbar\,m_1 m_2}{c^3 r^3}\,. \label{Newton}
\end{align}
The first term in the expression (\ref{Newton}) corresponds to the contribution
of gravitons, the second term corresponds to the contribution of photons and
the last one is the contribution of neutrinos, where $N_{\nu}$ is the number of
massless two-component neutrinos (we put $c=1$, $\hbar=1$ below).

In the case of interaction of scalar particles, all contributions logarithmic
in $q^2$ can be separated into two parts. The first one is the contributions in
which the coefficient before $\lnq$ is a function of $s=(p_1+p_2)^2$ with cuts
in the complex plane. These contributions can be generated by "box"$ $ diagrams
only (see Figs.~\ref{ff-interaction}~c,~d) that have unitary cut in the $s$- or
$u$-channel. These contributions, called in \cite{kk} irreducible, can be
calculated by the double dispersion relation in variables $s$ and $t=q^2$. The
distinctive characteristic of the contributions is that they are infrared
divergent (see expression (\ref{s3})) and, correspondingly, contain logarithms
where $q^2$ plays a role of an ultraviolet cutoff $\lnq/\lambda^2$. The second
part is the contributions that are entire rational functions of $s$. They arise
from the diagrams that have no unitary cut in the $s$- or $u$-channel and also
from the reducible part of the "box"$ $ diagrams. These contributions contain
logarithms with the transfer momentum as an infrared cutoff, for example,
logarithms that contain ultraviolet cutoff $\log\Lambda^2/\left\vert
q^2\right\vert$ or the masses of the interacting particles $\log
m_i^2/\left\vert q^2\right\vert$.

As shown in \cite{kk1}, the second part of the amplitude  can be represented as
a matrix element of an effective nonlocal operator of interaction of two
energy-momentum tensors (i.e., an effective graviton exchange) in the $q^2\to
0$ limit (but for arbitrary $s$):

\begin{align}
\hat{\mathcal{L}}_0 = \sqrt{-g} \,G^2\lnq
\left(\frac{23}{5}\,T_{\mu\nu}T^{\mu\nu} - \frac{31}{30} T^2 \right)
=\lnq\,\frac{\sqrt{-g}}{(8\pi)^2}\left(\frac{23}{5}\, R_{\mu\nu}R^{\mu\nu} -
\frac{31}{30} R^2\right),\label{operator}
\end{align}
where $T_{\mu\nu}, R_{\mu\nu}$ are the energy-momentum tensor of the matter and
the Ricci tensor, respectively. The main feature of the operator is that all
descriptions of the interacting particles appear only in the form of \textit{a
product} of the energy-momentum tensors, whereas the "propagator"$ $ of the
effective graviton is independent of the type of interacting
particles\footnote{Recall that here we are dealing with scalar particles
only.}. In contrast with the operator (\ref{operator}), the irreducible
contributions of the diagrams~\ref{ff-interaction}~c,~d contain
nonmultiplicative dependance on momenta and masses of the particles and cannot
be represented as operators like~(\ref{operator}).

On the basis of this operator, power quantum corrections to the Schwarzschild
metric were calculated in \cite{kk1}. Corrections to the interactions dependent
on internal angular momentum (spin) of a compound body were also considered.
There are three ways to calculate these corrections: first of all, it is
possible to calculate quantum corrections to the Kerr metric for rotating
compound body and to use covariant spin equation of motion in an external
gravitational field \cite{kp} as performed in \cite{kk1}; the second way is to
integrate velocity dependent interactions, also calculated in \cite{kk1}, over
the volumes of the interacting bodies; the last one is to put spin dependent
energy-momentum tensor to the operator (\ref{operator}) explicitly.

The purpose of this work is to calculate power quantum corrections to the
spin-dependent interactions for particles of spin $1/2$. It was shown in an
elementary way \cite{kk1} that spin-averaged contributions containing $q^2$ as
an infrared cutoff are exactly the same as (\ref{operator}). The irreducible
parts of the diagrams~\ref{ff-interaction}~c,~d averaged over spins coincide
with the scalar ones since the logarithm of the infrared cutoff $\log\lambda$
is cancelled by radiation of long-wave gravitons which is spin-independent.
However, it was noticed that the spin effects for rotating compound bodies with
scalar constituents can differ from the ones for particles of spin $1/2$. The
reason is that, apart from the operator (\ref{operator}), the leading
contribution to the spin effects can be generated by operators of higher order
in $q/m$. For example, first order in $q/m$ operators have the form:

\begin{eqnarray}
\hat{O}_{1} &=&\sqrt{-g}\,\lnq\,\frac{G}{16\pi }~R^{\mu \nu }{}_{\alpha \beta
;\mu
}\,t_{(a)}^{\alpha } t_{(b)}^{\beta } \left[ \sum\limits_{i}\frac{1}{4m_{i}}\bar{%
\psi}_{i}\left( -i\overleftarrow{D}_{\nu }\Sigma ^{ab}+\Sigma ^{ab}i%
\overrightarrow{D}_{\nu }\right) \psi _{i}\right]   \notag \\
&=&\sqrt{-g}\,\lnq\,\frac{G}{16\pi }\left( R^{\nu }{}_{\beta ;\alpha }-R^{\nu
}{}_{\alpha ;\beta }\right)t_{(a)}^{\alpha } t_{(b)}^{\beta } \left[
\sum\limits_{i}\frac{1}{4m_{i}}~\bar{\psi}_{i}\left( -i\overleftarrow{D}%
_{\nu }\Sigma ^{ab}+\Sigma ^{ab}i\overrightarrow{D}_{\nu }\right) \psi _{i}%
\right],  \label{operator1}\\
\hat{O}_{2} &=&\sqrt{-g}\,\lnq\,\frac{G^{2}}{2}~\left( T^{\nu }{}_{\beta
;\alpha }-T^{\nu }{}_{\alpha ;\beta }\right) t_{(a)}^{\alpha } t_{(b)}^{\beta }
\left[
\sum\limits_{i}\frac{1}{4m_{i}}~\bar{\psi}_{i}\left( -i\overleftarrow{D}%
_{\nu }\Sigma ^{ab}+\Sigma ^{ab}i\overrightarrow{D}_{\nu }\right) \psi _{i}%
\right],\label{operator2}
\end{eqnarray}
where left-hand-side and right-hand-side derivatives are defined as

\begin{align}
\overrightarrow{D}_{\alpha }=\overrightarrow{\partial }_{\alpha }-\frac{i}{2}%
\Sigma ^{ab}\gamma _{abc}~t_{\alpha }^{(c)}\,,\hspace{1cm} \overleftarrow{D}_{\alpha }=%
\overleftarrow{\partial }_{\alpha }+\frac{i}{2}\Sigma ^{ab}\gamma
_{abc}~t_{\alpha }^{(c)}\,.
\end{align}
Here $\Sigma ^{ab}=\frac{i}{2}~\sigma ^{ab}$ are the Lorentz group generators,
$\gamma _{abc}$ are the Ricci rotation coefficients, $t^\mu_{(a)}$ is a tetrad
of basis vectors that are specified by relation $t^\mu_{(a)}t_{\mu (b)}=\eta_{a
b}$, where $\eta_{ab}$ is the metric tensor of flat space-time. Taking into
account Einstein and Dirac equations in an external gravitational field

\begin{align}
R_{\mu \nu }=8\pi G\left( T_{\mu \nu }-\frac{1}{2}~g_{\mu \nu }T\right) ,%
\hspace{1cm}\left( i t_{b}^{\nu }\gamma ^{b}\overrightarrow{D}_{\nu }-m\right)
\psi =0\,,\label{ein}
\end{align}
note that the difference of the operators $\hat{O}_1$ and $\hat{O}_2$ is an
operator of second order in $q/m$

\begin{equation*}
\hat{O}_{1}-\hat{O}_{2}=\sqrt{-g}\,\lnq\,\frac{G^{2}}{16}\,T{}~\left( \sum\limits_{i}%
\frac{\Box }{m_{i}^{2}}~T_{i}\right)+\, total\: derivative
\end{equation*}
and makes zero contribution to the spin-dependent interactions.

Besides, contributions to the spin-spin interactions can be generated by
operators of the second order in $q/m$:

\begin{eqnarray}
\hat{O}_{3} &=&\sqrt{-g}\,\lnq\,D_{\mu }\left(
t^{\mu}_{(a)}~\bar{\psi}_{2}\gamma
^{5}\gamma ^{a}\psi _{2}\right) ~D_{\nu }\left( t^{\nu}_{(b)}~\bar{\psi}%
_{1}\gamma ^{5}\gamma ^{b}\psi _{1}\right) \label{operator3}\,, \\
\hat{O}_{4} &=&\sqrt{-g}\,\lnq\,D^{\mu }\left( t_{(a)}^{\nu
}~\bar{\psi}_{2}\gamma
^{5}\gamma ^{a}\psi _{2}\right) ~D_{\mu }\left( t_{\nu (b)}~\bar{\psi}%
_{1}\gamma ^{5}\gamma ^{b}\psi _{1}\right) \label{operator4}\,.
\end{eqnarray}

The plan of this paper is as follows: in Section 2 we describe useful Feynman
rules with effective vertices introduced in \cite{kk1} and also list possible
types of spin-dependent amplitudes with arbitrary coefficients (form factors).
The explicit expressions for all the form factors in the $t \to 0$ limit are
presented in Section 3. Results and conclusions are given in Sections 4 and~5.

\section{Effective vertices}

\begin{figure}[h]
\parbox{0.47\textwidth}{
\begin{center}
\includegraphics[width=6cm]{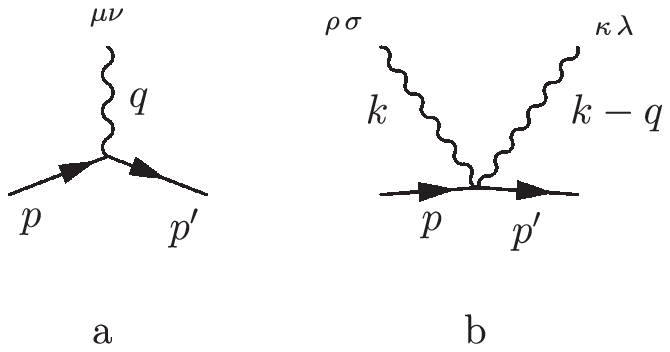}
\end{center}}
\parbox{0.47\textwidth}{
\begin{center}
\includegraphics[width=7.5cm]{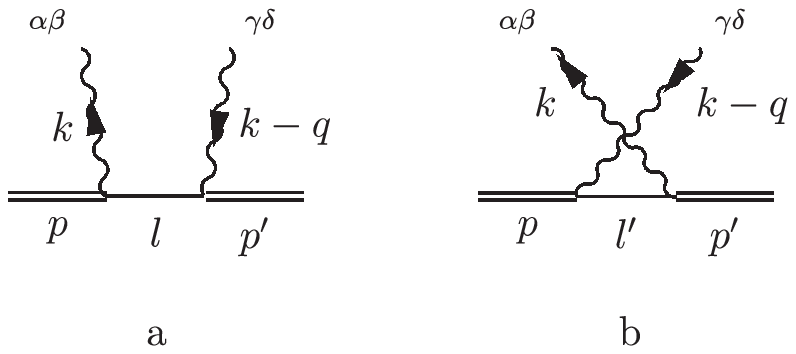}
\end{center}}
\parbox{0.47\textwidth}{
\caption{Vertices} \label{Vertices}}
\parbox{0.47\textwidth}{
\caption{Compton scattering amplitude} \label{Compton}}
\end{figure}
The single-graviton vertex for a spinor particle is (see Fig. \ref{Vertices}~a)

\begin{align}
V^{(1/2)}_{\mu \nu }=-\frac{i\kappa }{4}\,\bar{\psi}(p^{\prime })\left\{
I_{\mu \nu ,\,\alpha \beta }~P^{\alpha}\gamma ^{\beta }-\eta_{\mu \nu }\left( \hat{%
P}-2m\right) \right\} \psi(p),\label{vertex1}
\end{align}
here $\kappa=\sqrt{32\pi G}$, $P=p+p'$, $ \displaystyle I_{\mu \nu }{}^{\alpha
\beta }=\frac{1}{2}\left(\delta_\mu^\alpha \delta_\nu^\beta + \delta_\mu^\beta
\delta_\nu^\alpha\right) $ is a sort of a unit operator with the property
\[
I_{\mu\nu}{}^{\alpha\beta}t_{\alpha\beta}=t_{\mu\nu}
\]
for any symmetric tensor $t_{\alpha\beta}$. The last term in braces, expression
(\ref{vertex1}), proportional to the Minkowski metric tensor $\eta_{\mu\nu}$
vanishes on-mass-shell. It is convenient to exclude such terms from the
diagrams of Compton scattering amplitude containing a pole in the $s$- or
$u$-channel (see Fig. \ref{Compton}). This term cancels fermion propagator and
generates an additional contribution to the contact interaction of a mass-shell
fermion with two gravitons (see Fig. \ref{Vertices}~b). On rearrangement, the
double graviton vertex has a form:

\begin{align}
V_{\kappa \lambda ,~\rho \sigma }^{(1/2)}=i~\frac{3}{4}~\kappa ^{2}I_{\kappa
\lambda ,\alpha \delta }~I_{\beta}{}^{\delta}{}_{,\rho \sigma
}T_{(1/2)}^{\alpha \beta},\label{vertex2}
\end{align}
where $T^{(1/2)}_{\alpha \beta}$ is the energy-momentum tensor of the fermion
field

\begin{align}
T^{(1/2)}_{\mu \nu }=\frac{1}{2}\, I_{\mu \nu ,\,\alpha \beta }~P^{\alpha}
\bar{u}(p^{\prime })\gamma ^{\beta }\psi(p).
\end{align}
For further Feynman rules the reader is referred to the papers \cite{bb},
\cite{kk1}.

The rest of the pole diagrams (Fig.\ref{Compton}) is proportional to the
Compton scattering amplitude in quantum electrodynamics. For example, consider
diagram shown in Fig.~\ref{Compton}~a

\begin{align}
M_{\alpha\beta,\,\gamma\delta}=\frac{i\,\kappa^2}{16 (l^2-m^2)}\,
I_{\gamma\delta,\,\rho\sigma}\,I_{\alpha\beta,\,\phi\omega}\:
(p'+l)^\rho\,(p+l)^\phi
\left[\bar{u}\,(p')\,\gamma^\sigma\,(\hat{l}+m)\,\gamma^\omega\,u(p)\right].
\label{iline1}
\end{align}

It is useful to decompose the amplitude of scattering of photon by an electron
into independent spin structures:

\begin{multline}
M_{QED}^{\sigma\omega}=\bar{u}\,(p')\,\gamma^\sigma\,(\hat{l}+m)\,\gamma^\omega\,u(p)\\
=\frac{1}{2} \left[ (P-K)^{\sigma }\delta _{\rho }^{\omega }+(P-K)^{\omega
}\delta _{\rho }^{\sigma }-~q^{\sigma }\delta _{\rho }^{\omega }+\delta _{\rho
}^{\sigma }q^{\omega }+K_{\rho }~\eta^{\sigma \omega }\right] J^{\rho
}+\frac{i}{2}\,\epsilon ^{\sigma \omega \lambda \eta }K_{\lambda }S_{\eta
},\label{iline2}
\end{multline}
here $P=2p-q,\,K=2k-q$, and $J^\mu$, $S^\mu$ are the vector current and spin
four-vector (axial current), respectively:

\begin{eqnarray}
J^\mu=\bar{u}\,(p')\,\gamma^\mu\,u(p)\,, &\quad S^\mu=\bar{u}\,(p')\,\gamma_5\,
\gamma^\mu\,u(p)\,, &\quad \gamma _{5}=\left(
\begin{array}{cc}
0 & -1 \\
-1 & 0
\end{array}%
\right).
\end{eqnarray}
Diagram shown in Fig.~\ref{Compton}~b can be considered in a similar manner.
Vector current $J^\mu$ can be also decomposed into a convective part and a part
containing spin explicitly:

\begin{align}
\left( 1-\frac{q^{2}}{4m^{2}}\right) ~J^{\mu }=\frac{P^{\mu
}}{2m}~\bar{u}(p^{\prime})u(p) +\frac{i}{4m^2}\,\epsilon ^{\mu \gamma \alpha
\delta }q_{\gamma}P_{\alpha }S_{\delta }.
\end{align}
Therefore, all vertices and lines can be decomposed into two structure
including bispinor amplitudes of a fermion, namely, $\bar{u}(p^{\prime})u(p)$
and $S^\mu$. Thus, matrix element of the fermion-fermion scattering can be
separated into three groups. The first one does not include $S^{1,2}_\mu$
explicitly:

\begin{align}
\mathcal{M}_{N}(q)=\mathcal{F}_N(s)\, G^{2} m_1 m_2~\left(
\bar{u}_{1}u_{1}\right)\left( \bar{u}_{2}u_{2}\right)\lnq\label{Namp},
\end{align}
the second one contains the terms which are proportional to the first power of
$S^{1,2}_\mu$:

\begin{align}
\mathcal{M}_{SO}(q)=i \mathcal{F}_{SO}(s)\,G^{2}\, \epsilon _{\alpha \,\beta
\,\gamma \,\delta }\,q^{\alpha }\,P_{1}^{\beta }\,P_{2}^{\gamma }\left(
\frac{S_{2}^{\delta }}{m_{2}}\left( \bar{u}_{1}u_{1}\right) +
\frac{S_{1}^{\delta }}{m_{1}} \left(
\bar{u}_{2}u_{2}\right)\right)\lnq\label{SOamp},
\end{align}
and the last part contains terms quadratic in spin:

\begin{align}
\mathcal{M}_{SS}=G^{2} \left( \mathcal{F}^{(1)}_{SS}(s)\,(q\cdot
S_{1})\,(q\cdot S_{2})+\mathcal{F}^{(2)}_{SS}(s) \,t\,(S_{1}\cdot S_{2})\right)
\lnq \label{SSamp}.
\end{align}

In the first Born approximation corrections to a potential are given by a
Fourier transform of nonrelativistic amplitude

\begin{align}
U(\mathbf{r})=\int \frac{d^{3}\mathbf{q}}{(2\pi )^{3}}~\mathcal{A}(\mathbf{q}%
)~e^{-i\mathbf{q}\cdot \mathbf{r}}=-\int \frac{d^{3}\mathbf{q}}{(2\pi )^{3}}%
~e^{-i\mathbf{q}\cdot \mathbf{r}}\lim_{s\rightarrow
(m_{1}+m_{2})^{2}}\lim_{t\rightarrow 0}~\frac{\mathcal{M}}{\prod\limits_{i}\sqrt{2E_{i}%
}}\,.
\end{align}

Since we are interested in amplitudes in the $t\to 0$ limit, the form factors
$\mathcal{F}_{i}$ are the functions of the variable $s$. It is convenient to
introduce a new variable

\begin{align}\gamm =(s-m_{1}^{2}-m_{2}^{2})/(2 m_{1} m_{2})
\end{align}
which corresponds to the $\gamma$-factor of one of the interacting particles in the
limit when the mass of the other goes to infinity.

\section{Matrix elements}
\begin{figure}[ht]
\parbox{0.4\textwidth}{
\begin{center}
\includegraphics[width=4cm]{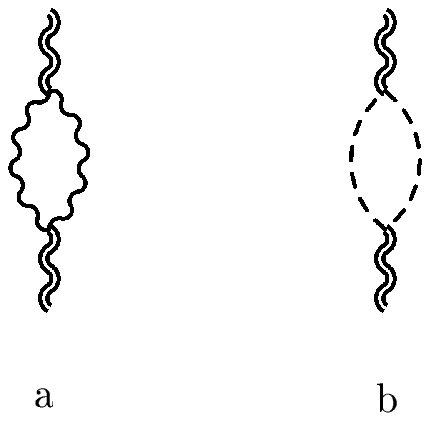}
\end{center}}
\hfill
\parbox{0.4\textwidth}{
\begin{center}
\includegraphics[width=5.5cm]{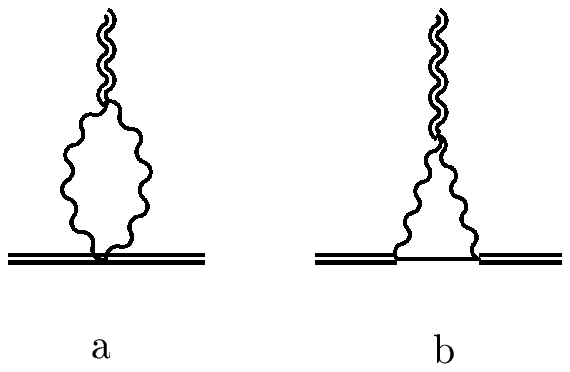}
\end{center}}
\parbox{0.4\textwidth}{
\caption{Graviton-graviton interaction} \label{gg-interaction}} \hfill
\parbox{0.4\textwidth}{
\caption{Graviton-fermion interaction} \label{gf-interaction}}
\end{figure}

We start the discussion of loops with the graviton-graviton interaction shown
in Fig.~\ref{gg-interaction}. It is easy to calculate this contribution using
the effective operator derived in \cite{hoo}

\begin{align}
\mathcal{L}_{gg}=-\lnq\,\frac{\sqrt{-g}}{(8 \pi)^{2}}\,\left(
\frac{21}{15}\,R^{\mu \nu }R_{\mu \nu }+\frac{1}{30}\,R^{2}\right).\label{gg}
\end{align}%
By substituting the Ricci tensor from the Einstein equations (\ref{ein}) one
gets the following matrix elements:

\begin{align}
\mathcal{F}_{N}(\gamm) & =-\frac{1}{15}~\left(  42\gamm^{2}+1\right)
,\hspace{1cm}\mathcal{F}_{SO}(\gamm)=-\frac{7}{20}~\gamm\,,\\
\mathcal{F}_{SS}^{(1)}(\gamm) & =-\mathcal{F}_{SS}^{(2)}(\gamm)=\frac{7}%
{20}\left(  1-2\gamm^{2}\right)  .
\end{align}

Diagrams containing the graviton-fermion interaction (see
Fig.~\ref{gf-interaction}) make the following contribution to the matrix
elements:

\begin{align}
\mathcal{F}_{N}(\gamm)=2\left(  6\gamm^{2}+1\right)  ,\hspace{.5cm}%
\mathcal{F}_{SO}(\gamm)=\frac{\gamm}{2},\hspace{.5cm}\mathcal{F}_{SS}%
^{(1)}(\gamm)=-\mathcal{F}_{SS}^{(2)}(\gamm)=\frac{1}{2}\left(  1-2\gamm
^{2}\right)  .
\end{align}
These corrections coincide with the corresponding form factor corrections
calculated in \cite{bbo}. By analogy with (\ref{gg}) the contribution of the
diagrams including graviton-fermion interaction (Fig.~\ref{gf-interaction}) can
be represented as an effective operator:

\begin{eqnarray}
\mathcal{L}_{gf}=\sqrt{-g}\,\lnq\,\frac{G}{4 \pi}\,\left( 3 R^{\mu\nu}
T_{\mu\nu}-2 R T\right)- 8\,\hat{O}_1\,,\label{gf}
\end{eqnarray}
where the operator $\hat{O}_1$ is defined by (\ref{operator1}).

\begin{figure}[ht]
\begin{center}
\includegraphics[width=4.5cm]{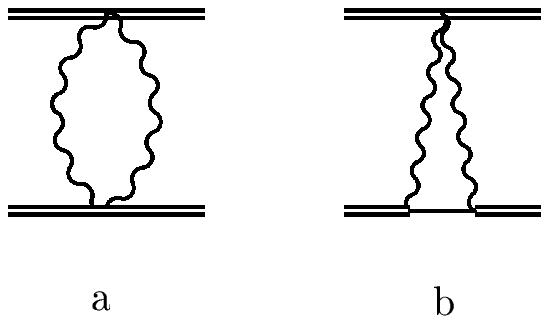}\hspace{.3cm}
\includegraphics[width=4.5cm]{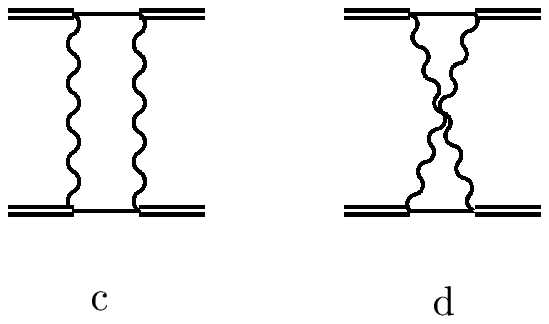}
\end{center}
\caption{Fermion-fermion interaction}\label{ff-interaction}
\end{figure}

The rest of the diagrams (Fig.~\ref{ff-interaction}) corresponds to the
fermion-fermion interaction. The sum of the diagrams
Figs.~\ref{ff-interaction}~a,~b makes the following contribution to the matrix
elements:
\begin{align}
\mathcal{F}_{N}(\gamm)=-3\left(  5\gamm^{2}-1\right)  ,\hspace
{.5cm}\mathcal{F}_{SO}(\gamm)=-\frac{15}{16}~\gamm,\hspace{.5cm}%
\mathcal{F}_{SS}^{(1)}(\gamm)=-\mathcal{F}_{SS}^{(2)}(\gamm)=0.\label{ff_e1}
\end{align}
The sum of these diagrams does not contribute to the spin-spin interaction. The
corresponding effective operator is

\begin{eqnarray}
\mathcal{L}^{\mathrm(ab)}_{ff}=- \sqrt{-g}\,\lnq\,G^2\left( \frac{15}{2}\,
T^{\mu\nu} T_{\mu\nu}-\frac{3}{2}\,T^2 \right)+\frac{15}{2}\, \hat{O}_2
\label{ff-1}\,,
\end{eqnarray}
where the operator $\hat{O}_2$ is defined by (\ref{operator2}).

There are a number of points to be made before considering the contribution of
the diagrams Figs.~\ref{ff-interaction}~c,~d. As stated above, the analysis of
these diagrams is complicated by the presence of infrared divergencies. Aside
from the integrals (\ref{int_b})-(\ref{int_e}) containing $q^2$ as an infrared
cutoff, there are contributions generated by integrals (\ref{s3}), (\ref{u3})
containing $q^2$ as an ultraviolet cutoff. There is a unitary cut in the $s$-
or $u$- channel in the last-mentioned integrals, and consequently, there is a
nonpolynomial dependence on $s$. We present the results for the reducible and
irreducible parts for each type of the matrix elements separately. However, as
we shall see later, the reducible part also contains a nonpolynomial dependence
on $s$. The reason is a singularity in the $\Delta\rightarrow 0$ limit in
vector and tensor integrals of the $K^{\mathrm{(s)}}_{\mu}$ type (see
expression (\ref{sv3}) in Appendix) where

\begin{align}
& \Delta = \frac{1}{4 s}\,\big((P_1 P_2)^2-P_1^2 P_2^2\big), \qquad
P_1=p_1+p_1',\quad P_2=p_2+p_2'\,.
\end{align}
In the center of mass frame the variable $\Delta$ takes the form:

\begin{align}
\Delta = 4 |\vec{p}\,|^2 \cos^2\theta/2.
\end{align}
Terms proportional to $\log(\sin\theta/2)/(\cos^2\theta/2)$ are typical for the
diagrams of the sort of Figs.~\ref{ff-interaction}~c,~d. For example, they
occur in the calculation of the polarization vector when an electron is
scattered by the Coulomb field in the second Born approximation. Terms
proportional to $1/\Delta$ cancel in the spin independent part of the matrix
elements (\ref{Namp}) in the reducible and irreducible parts independently.
However, in the spin dependent terms (\ref{SOamp}), (\ref{SSamp}) $1/\Delta$
cancels only in the sum of the irreducible and reducible parts in the
nonrelativistic limit.

Taking into account all mentioned above, we write out the matrix elements for
the diagrams \ref{ff-interaction}~c,~d. The reducible parts are

\begin{align}
\mathcal{F}_{N}(\gamm)& = -7+15~\gamm^{2}\,,& \mathcal{F}_{SO}(\gamm)&
=\frac{47}{16}~\gamm+\frac{7}{8}\frac{\gamm}{\gamm
^{2}-1}  \label{PP&S_Red}\,,\\
\mathcal{F}_{SS}^{(1)}(\gamm)  & =-4(1-2~\gamm^{2})+\frac{31}{12}+\frac{1}{2}\frac{1}{\gamm^{2}%
-1} \,,&
\mathcal{F}_{SS}^{(2)}(\gamm)  & =4(1-2~\gamm^{2})-\frac{77}{24}-\frac{1}{2}\frac{1}{\gamm^{2}%
-1}\label{SS_Red}\,.%
\end{align}
Since in the expressions (\ref{PP&S_Red}), (\ref{SS_Red}) we take the limit
$t\to 0$, the singularity $1/\Delta$ turns to the singularity $1/(\gamm^2-1)$.
Terms not proportional to $1/(\gamm^2-1)$ can be again represented in the form
of an effective operator

\begin{eqnarray}
\mathcal{L}^{\mathrm(cd)}_{ff}= \sqrt{-g}\,\lnq\,G^2\left( \frac{15}{2}\,
T^{\mu\nu} T_{\mu\nu}-\frac{7}{2}\,T^2 \right)+\frac{17}{2}\, \hat{O}_2
+\frac{31}{12}\,\hat{O}_3-\frac{77}{24}\,\hat{O}_4\label{ff-2}\,.
\end{eqnarray}

The irreducible parts are

\begin{align}
\mathcal{F}_{N}(\gamm) & =-16\,\gamm\,(2\gamm^{2}-1)\,\mathcal{W}_{A}(\gamm)+\frac{2}%
{3}~(1-2\gamm^{2})^{2}\,\mathcal{W}_{S}(\gamm)\label{PP_Irred}\,,\\
\mathcal{F}_{SO}(\gamm)& =-\left(  4\gamm^{2}+\frac{1}{2}+\frac{7}{8}\frac{1}%
{\gamm^{2}-1}\right)  \mathcal{W}_{A}(\gamm)+\frac{\gamm}{6}\,\left( 2\gamm^2-1
\right)  \mathcal{W}_{S}(\gamm)\label{PS_Irred}\,,\\
\mathcal{F}_{SS}^{(1)}(\gamm)  & =\left(  2\gamm-8\gamm^{3}-\frac{1}{2}%
\frac{\gamm}{\gamm^{2}-1}\right)  \mathcal{W}_{A}(\gamm)+\frac{1}%
{6}~\left(  1-2\gamm^{2}\right)  ^{2}\mathcal{W}_{S}(\gamm)\,,\label{SS_Irred_1}\\
\mathcal{F}_{SS}^{(2)}(\gamm)  & =\left(  -2\gamm+8\gamm^{3}+\frac{1}{2}%
\frac{\gamm}{\gamm^{2}-1}\right)  \mathcal{W}_{A}(\gamm)+\left(  -\frac
{1}{6}~\left(  2\gamm^{2}-1\right)  ^{2}+\frac{1}{12}\right)  \mathcal{W}%
_{S}(\gamm)\,, \label{SS_Irred_2}
\end{align}
where functions $\mathcal{W}_{S}(\gamm),\,\mathcal{W}_{A}(\gamm)$ are defined
in Appendix by Eqs.~(\ref{lws}), (\ref{lwa}). As noted above, the terms
singular in the $\gamm\to 1$ limit cancel in the sum of the reducible
(\ref{PP&S_Red}), (\ref{SS_Red}) and irreducible
(\ref{PS_Irred})-(\ref{SS_Irred_2}) parts

\begin{align}
\mathcal{F}_{SO}(1)=-\frac{13}{16}\,, \qquad
\mathcal{F}^{(1)}_{SS}(1)=\frac{7}{12}\,, \qquad
\mathcal{F}^{(2)}_{SS}(1)=-\frac{9}{8}\,.
\end{align}

\section{Quantum corrections to spin effects}

Summing up all the contributions we get the following scattering amplitudes.
The terms not containing $S^{1,2}_\mu$ explicitly are

\begin{equation}
\mathcal{A}_{N}(q)=\frac{41}{5}\,G^{2}{m_{1}}\,{m_{2}} ~\frac{\left(
\bar{u}_{1}u_{1}\right) }{2m_{1}}\frac{\left( \bar{u}_{2}u_{2}\right)
}{2m_{2}}\,\lnq\,.
\end{equation}
The terms proportional to the first power of $S^{1,2}_\mu$ are

\begin{equation}
\mathcal{A}_{SO}(q)=\frac{8}{5}\,i~G^{2}\,\epsilon _{\alpha \,\beta \,\gamma
\,\delta }\,q^{\alpha }\,P_{1}^{\beta }\,P_{2}^{\gamma }\left(
\frac{S_{2}^{\delta }}{2m_{2}^{2}}\,\frac{\left(
\bar{u}_{1}u_{1}\right) }{2m_{1}}{+}\frac{S_{1}^{\delta }}{2m_{1}^{2}}\frac{%
\left( \bar{u}_{2}u_{2}\right) }{2m_{2}}\right)\lnq\,.
\end{equation}
At last, the part containing terms quadratic in spins is

\begin{equation}
\mathcal{A}_{SS}=\frac{G^{2}}{4m_{1}m_{2}}
\left( \frac{4}{15}\,(q\cdot S_{1})\,(q\cdot S_{2})+\frac{11}{40}%
\,t\,(S_{1}\cdot S_{2})\right)\lnq\,.
\end{equation}

Using the nonrelativistic expansions of the bispinor amplitudes of the
scattering particles

\begin{align}
& \frac{1}{2m_{1,2}}~\bar{u}(p_{1,2}\mp q)~u(p_{1,2})
= 1\pm~\frac{i}{2m_{1,2}^{2}}~\epsilon _{ijk}\mathbf{q}%
^{i}\mathbf{p}_{1,2}^{j}\mathbf{s}_{1,2}^{k}+\ldots\\
& \frac{1}{2m_{1,2}}~S_{1,2}^{\delta }=\frac{1}{2m_{1,2}}~\bar{u}(p_{1,2}\mp
q)\gamma_{5}
\gamma ^{\delta }\bar{u}(p_{1,2})\approx ~\left( 0,2\mathbf{s}_{1,2}\right)\,\\
& \frac{i}{2m_{1,2}^{2}}~\epsilon _{\alpha \,\beta \,\gamma \,\delta
}\,q^{\alpha }\,P_{1}^{\beta }\,P_{2}^{\gamma }S_{1,2}^{\delta } =\pm
8i~\epsilon _{n\,k\,m}\,\mathbf{q}^{n}\mathbf{p}_{2,1}^{k}
\mathbf{s}_{1,2}^{m}\mp 8i~\frac{m_{2,1}}{m_{1,2}}~\epsilon
_{nk\,m}\,\mathbf{q}^{n} \mathbf{p}_{1,2}^{k}\mathbf{s}_{1,2}^{m},
\end{align}
yields the amplitudes in the momentum space:

\begin{align}
\mathcal{A}_{N}(\mathbf{q})& = \frac{41}{5}\,G^{2}{m_{1}}\,{m_{2}}\log
\mathbf{q}^{2}
\label{Newton-corr}\\
\mathcal{A}_{SO}(\mathbf{q})
&= \frac{87}{10}G^{2}\,\left( -~i~\frac{m_{2}}{m_{1}}~\epsilon _{ijk}\mathbf{q%
}^{i}\mathbf{p}_{1}^{j}\mathbf{s}_{1}^{k}+i~\frac{m_{1}}{m_{2}}~\epsilon
_{ijk}\mathbf{q}^{i}\mathbf{p}_{2}^{j}\mathbf{s}_{2}^{k}\right) \log \mathbf{%
q}^{2}\,, \\
&+ \frac{64}{5}~G^{2}\,\left( i~\epsilon _{n\,k\,m}\,\mathbf{q}^{n}\mathbf{%
p}_{2}^{k}\mathbf{s}_{1}^{m}-i~\epsilon _{nk\,m}\,\mathbf{q}^{n}\mathbf{p}%
_{1}^{k}\mathbf{s}_{2}^{m}\right) \log \mathbf{q}^{2}\,, \label{SO-corr}\\
\mathcal{A}_{SS}(\mathbf{q})&= G^{2}\left( \frac{16}{15}\,(\mathbf{q}\cdot \mathbf{s}%
_{1})\,(\mathbf{q}\cdot \mathbf{s}_{2})+\frac{11}{10}\,\mathbf{q}^{2}(%
\mathbf{s}_{1}\cdot \mathbf{s}_{2})\right) ~\log \mathbf{q}^{2}\,.
\label{SS-corr}
\end{align}

With the expressions (\ref{Fourier}) from Appendix we obtain the corrections in
the coordinate space. First of all, according to \cite{bb}, \cite{kk1}, the
power quantum correction to the Newton law is the same as for scalar particles
(\ref{Newton})

\begin{equation}
U_{N}=-\frac{41}{10}\frac{\hbar\, G^{2}{m_{1}}\,{m_{2}}}{c^3 \pi r^{3}}\,.
\end{equation}

The quantum correction to the gravitational spin-orbit interaction is

\begin{align}
U_{SO} & =\frac{261}{20}\frac{\hbar\, G^{2}}{c^5 \pi r^{5}}\left( \frac{m_{2}}{m_{1}}~%
\mathbf{s}_{1}+~\frac{m_{1}}{m_{2}}~\mathbf{s}_{2}\right)\cdot \mathbf{l}\,,\label{so}\\
\mathbf{l}^{i} & =
\epsilon^{ijk}~(\mathbf{r}_1-\mathbf{r}_2)_{j}~\mathbf{p}_{k}\,,
\end{align}
where $\mathbf{p}=\mathbf{p}_1=-\mathbf{p}_2$ is the relative momentum.

The quantum correction to the Lense-Thirring effect is

\begin{equation}
U_{LT}=\frac{96}{5}\frac{\hbar\, G^{2}}{c^5 \pi
r^{5}}\left(\mathbf{s}_{1}+\mathbf{s}_{2}\right)\cdot \mathbf{l}\,.\label{lt}
\end{equation}

The quantum correction to the interaction of the two internal angular momenta
$\mathbf{s}_i$, i.e., to the gravitational spin-spin interaction is

\begin{equation}
U_{SS}=\frac{\hbar\, G^{2}}{c^5 \pi r^{5}}\left( 8\,(\mathbf{n}\cdot \mathbf{s%
}_{1})\,(\mathbf{n}\cdot \mathbf{s}_{2})+\frac{5}{2}\,(\mathbf{s}_{1}\cdot
\mathbf{s}_{2})\right) .\label{ss}
\end{equation}

\section{Discussion and conclusions}
Hence, as expected, the quantum corrections to the classical Hamiltonian
(\ref{so}), (\ref{lt}), (\ref{ss}) for a spinor particle differ from the full
sum of the similar corrections obtained in \cite{kk1} for a rotating compound
body with scalar constituents.

This difference is associated with the irreducible part as well as with the
reducible one that can be represented in the form of an effective operator (see
(\ref{operator}), (\ref{gg}), (\ref{gf}), (\ref{ff-1}), (\ref{ff-2}))

\begin{align}
\mathcal{L}=\mathcal{L}_0 -8\hat{O}_1 + 16
\hat{O}_2+\frac{31}{12}\,\hat{O}_3-\frac{77}{24}\,\hat{O}_4\,,\label{e_2}
\end{align}
where the new operators $\hat{O}_i$ take part.

Following are several points concerning the operator (\ref{e_2}). As
demonstrated in \cite{kk1}, after averaging over the spins, the Compton
amplitude of fermion-graviton scattering and the scalar-graviton one coincide
with the required accuracy. In particular, it means that the effects linear in
spin (see expressions (\ref{so}), (\ref{lt})) for fermion-fermion and
fermion-scalar scattering are the same. However, we see that the analysis is
complicated by the singularity $1/\Delta$. Therefore, it is not unreasonable to
make sure that the separation into the reducible and the irreducible parts is
the same as in Eqs.(\ref{ff-1}), (\ref{PP&S_Red}), (\ref{ff-2}),
(\ref{PP_Irred}) for the case of fermion scattering by a scalar particle
(evidently, there is only a contribution of one of the particles under the
summation sign in the operators (\ref{ff-2}), (\ref{PP_Irred}) in this case).
We use the effective vertices derived in \cite{kk1} for scalar particles. The
diagrams Figs. \ref{ff-interaction}~a,~b make the following contribution to the
spin effects:

\begin{align}
\mathcal{F}_{N}(\gamm)=5-16\gamm^{2},\hspace
{1cm}\mathcal{F}_{SO}(\gamm)=-\frac{3}{4}~\gamm\,.\label{scalar_e1}%
\end{align}
The reducible part of the diagrams Figs. \ref{ff-interaction}~c,~d yields

\begin{align}
\mathcal{F}_{N}~(\gamm )=-9+16~\gamm ^{2},\hspace {1cm} \mathcal{F}_{SO}~(\gamm
)=\frac{11}{4}~\gamm +\frac{7}{8}\frac{\gamm }{\gamm ^{2}-1}.\label{scalar_e2}
\end{align}
Note that the sum of (\ref{scalar_e1}) and (\ref{scalar_e2}) agree with the sum
of corresponding expressions (\ref{ff_e1}) and (\ref{PP&S_Red}). The
irreducible part is also equal to (\ref{PS_Irred}). Therefore the singularities
$1/(\gamm^2-1)$ cancel in the same way as in the case of fermion-fermion
scattering.

Thus, the effective operator (\ref{e_2}) is universal for scattering of scalar
particles and particles of spin $1/2$.

Being interested in the quantum corrections to spin effects at the leading
order in $1/c$ (where $c$ is the speed of light), we considered only the matrix
elements like (\ref{Namp})-(\ref{SSamp}). Formally, it is possible to consider
one more matrix element at the same order in $q/m_i$

\begin{align}
\mathcal{M}(s,t)=\mathcal{F}(s,t)~t~\frac{\left( P_{1}S_{2}\right) }{m_{2}}%
\frac{\left( P_{2}S_{1}\right) }{m_{1}},\label{e_4}
\end{align}
that however does not contribute to the spin effects at the leading order in
$1/c$. Nevertheless, it is possible to make sure that the contribution of the
diagrams Figs. \ref{gg-interaction}, \ref{gf-interaction},
\ref{ff-interaction}~a,~b and the reducible part of the diagram Figs.
\ref{ff-interaction}~c,~d to the the form factor $\mathcal{F}(s,t)$ can also be
described by the operator (\ref{e_2}). This fact is trivial for the diagrams
presented in Fig. \ref{gg-interaction} since their contribution originally
contains the energy-momentum tensors of interacting particles. Contribution of
the diagrams Fig. \ref{gf-interaction} is

\begin{align}
\mathcal{F}(\gamm )=-\frac{\gamm }{4}\,,
\end{align}
that corresponds to the contribution of the operator (\ref{gf}). The diagrams
Figs.~\ref{ff-interaction}~a,~b yield

\begin{align}
\mathcal{F}(\gamm )=0\,,
\end{align}
that is also equal to the contribution of the corresponding operator
(\ref{ff-1}). The reducible contribution of the diagrams
Figs.(\ref{ff-interaction})c,~d is

\begin{align}
\mathcal{F}_{\mathrm{red}}(\gamm )=2\gamm -\frac{33\gamm }{16~(1-\gamm
^{2})}+\frac{\gamm }{4(1-\gamm ^{2})^{2}}.\label{e_3}
\end{align}

Note that the terms not containing $1/(1-\gamm^2)$ also correspond to the
contribution of the operator (\ref{ff-2}). Hence, the operator (\ref{e_2})
describes all the contributions to the matrix element (\ref{e_4}) correctly.
Terms in the expression (\ref{e_3}) singular in the $\gamm\to 1$ limit cancel
with the irreducible part of the diagrams Figs.~\ref{ff-interaction}c,~d in the
same manner as for the matrix elements (\ref{SOamp}), (\ref{SSamp}):

\begin{align}
\mathcal{F}_{\mathrm{irred}}(\gamm )=\left( \frac{\gamm ^{3}}{6}+\frac{%
\gamm }{48~(1-\gamm ^{2})}\right) \mathcal{W}_{S}+\left( -2(1+\gamm ^{2})+%
\frac{9}{4(1-\gamm ^{2})}-\frac{1}{4(1-\gamm ^{2})^{2}}\right)
\mathcal{W}_{A}\,,
\end{align}
\begin{align}
 \lim_{\gamm \rightarrow 1}\left( \mathcal{F}_{\mathrm{red}}(\gamm )+%
\mathcal{F}_{\mathrm{irred}}(\gamm )\right) =-\frac{19}{40}\,.
\end{align}

\vspace{.5cm}

I would like to thank I.B. Khriplovich and A.A. Pomeransky for their helpful
comments and discussions. The investigation was supported by the Russian
Foundation for Basic Research through Grant No. 05-02-16627-a and by Institute
$\mathrm{f\ddot{u}r}$ Theoretische Physik-II Ruhr-$\mathrm{Universit\ddot{a}t}$
through Graduiertenkolleg "Physik der Elementarteilchen an Beschleunigern und
im Universum".

\section*{Appendix}
To calculate the correction to spin-dependent interactions, it is needed to
consider the following Fourier transforms:

\begin{align}
& \int \log \mathbf{q}^{2}e^{-i\mathbf{q}\cdot \mathbf{r}}\frac{d^{3}\mathbf{q}%
}{(2\pi )^{3}} = -\frac{1}{2\pi r^{3}}\,, & &
\int \left( \mathbf{q}_{i}\mathbf{q}_{j}\,\log \mathbf{q}^{2}\right)
~e^{-i\mathbf{q}\cdot \mathbf{r}}\frac{d^{3}\mathbf{q}}{(2\pi )^{3}} =
\frac{3}{2\pi r^{5}}\left( 5~\mathbf{n}_{i}~\mathbf{n}_{j}-\delta _{ij}\right)\,, \nonumber\\
& \int \left( -i\mathbf{q}_i~\log \mathbf{q}^{2}\right) ~e^{-i\mathbf{q}\cdot
\mathbf{r}}\frac{d^{3}\mathbf{q}}{(2\pi )^{3}} = \frac{3}{2\pi
}\frac{\mathbf{r}_i}{r^{5}}\,,   & & \int \left( \mathbf{q}^{2}\log
\mathbf{q}^{2}\right) ~e^{-i\mathbf{q}\cdot
\mathbf{r}}\frac{d^{3}\mathbf{q}}{(2\pi )^{3}} =\frac{3}{\pi r^{5}}\,.
\label{Fourier}
\end{align}

According to the choice $\langle1(p_1-q),2(p_2+q)\vert1(p_1),2(p_2)\rangle$ of
kinematics and the sign in the exponent of the Fourier transforms, the vector
$\mathbf{r}$ is directed from the second body to the first one, i.e.,
$\mathbf{r}=\mathbf{r}_1-\mathbf{r}_2$.

Logarithmic parts of the integrals occurring are given below, using the
notation $L=-\frac{i}{(4\pi)^2}\,\lnq$ :

\begin{eqnarray}
& I=\intk &\frac{1}{k^2 (k-q)^2}= L+\ldots\label{int_b}\\
& I_\mu=\intk &\frac{k_\mu}{k^2 (k-q)^2}=\frac{q_\mu}{2} L+\ldots\\
& I_{\mu\nu}=\intk &\frac{k_\mu k_\nu}{k^2 (k-q)^2}
=\frac{1}{3}\left(q_\mu q_\nu -\frac{q^2}{4}\delta_{\mu\nu}\right) L+\ldots\\
& J=\intk &\frac{1}{k^2 (k-q)^2 \left((p-k)^2-m^2\right)}=\frac{L}{2 m^2} +\ldots\label{int_tr}\\
& J_\mu=\intk &\frac{k_\mu}{k^2 (k-q)^2 \left((p-k)^2-m^2\right)}
=-\frac{{p}_{\mu}-q_\mu}{2 m^2}\,L+\ldots\\
& J_{\mu\nu}=\intk &\frac{k_\mu k_\nu}{k^2 (k-q)^2\left((p-k)^2-m^2\right)}=\\
&&\frac{L}{4 m^2}\left\{\frac{q^2}{m^2}\, p_\mu p_\nu - (p_\mu q_\nu + p_\nu
q_\mu)+2q_\mu q_\nu- \frac{q^2}{2}\,g_{\mu\nu}\right\}+\ldots \label{int_e}
\end{eqnarray}

For the diagrams Fig.~\ref{ff-interaction}~c,~d to be calculated, it is
necessary to consider integrals with four Feynman propagators with the infrared
divergency regularized by the "mass"$ $ of graviton $\la$. The imaginary part
of the diagram Fig.\ref{ff-interaction}~c will be omitted because it does not
contribute to the potential of the two-body interaction. It corresponds to the
second Born approximation. The real part of the integral for the diagram
Fig.\ref{ff-interaction}~c is

\begin{align}
K^{\mathrm{(s)}}  & =\int\frac{d^{D}k}{(2\pi)^{D}}\frac{1}{\left(
k^{2}+i0\right) \left( (q-k)^{2}+i0\right)  \left(
(p_{1}-k)^{2}-m_{1}^{2}+i0\right)
\left(  (p_{2}+k)^{2}-m_{2}^{2}+i0\right)  }\nonumber\\
& =-\frac{i}{(4\pi)^{2}}~\frac{1}{m_{1}m_{2}~t}\log\left\vert \frac{t}%
{\lambda^{2}}\right\vert \frac{2~m_{1}m_{2}}{\sqrt{\left(  s-m_{+}^{2}\right)
\left(  s-m_{-}^{2}\right)  }}\log\frac{\sqrt{s-m_{-}^{2}}-\sqrt{s-m_{+}^{2}}%
}{\sqrt{s-m_{-}^{2}}+\sqrt{s-m_{+}^{2}}}+\ldots\nonumber \\
& =-\frac{i}{(4\pi)^{2}}~\frac{1}%
{m_{1}m_{2}~t}\log\left\vert \frac{t}{\lambda^{2}}\right\vert ~\mathcal{K}%
_{s}(s)+\ldots\,,\label{s3}
\end{align}
where notations $m_{+}=m_1+m_2,\, m_{-}=m_1-m_2$ were introduced, and
dependence on the variable $s$ denoted as the function $\mathcal{K} _{s}(s)$.
To simplify expressions it is convenient to introduce a new variable $\gamm
=(s-m_{1}^{2}-m_{2}^{2})/(2 m_{1} m_{2})$.

\begin{align}
\mathcal{K}_{s}(s)  & =\frac{1}{\sqrt{\gamm^{2}-1}}~\log\left(
\gamm-\sqrt{\gamm^{2}-1}\right)\,.
\end{align}
In the diagram Fig.\ref{ff-interaction}~d the same integral can be derived by
 analytic continuation of all the expressions from the $s$- to the
$u$-channel ($t=q^2, u=2 m^2_1+2 m^2_2-s-t$):

\begin{align}
\mathcal{K}_{u}(u)=-\mathcal{K}_{s}(2 m^2_1+2 m^2_2-s-t)=\frac{2~m_{1}m_{2}}{\sqrt{\left(  m_{+}^{2}-u\right)
\left(  m_{-}^{2}-u\right)  }}\log\frac{\sqrt{m_{+}^{2}-u}+\sqrt{m_{-}^{2}-u}%
}{\sqrt{m_{+}^{2}-u}-\sqrt{m_{-}^{2}-u}}\,.\label{u3}
\end{align}
In the limit when the momentum transfer $q^\mu$ goes to zero the function
$\mathcal{K}_{u}(u)$ can be expressed as a function of $\mathcal{K}_{s}(s)$ up
to terms linear in $t$:

\begin{align}
\mathcal{K}_{u}(u) \approx-\mathcal{K}_{s}(s)+\frac{t}{2m_{1}m_{2}}\frac
{\gamm~\mathcal{K}_{s}(s)+1}{\gamm^{2}-1}\,.%
\end{align}
In the course of the calculation, it is convenient to express amplitudes as a
half-sum and half-difference of the functions $\mathcal{K}_{s}(s)$ and
$\mathcal{K}_{u}(u)$:

\begin{align}
\mathcal{W}_{S}(\gamm)  & =-\lim_{t\rightarrow0}\frac{6m_{1}m_{2}}{t}\frac
{1}{2}\left(  \mathcal{K}_{s}(s)+\mathcal{K}_{u}(u)\right)  =-3~\frac
{\gamm~\mathcal{K}_{s}(s)+1}{\gamm^{2}-1} \label{lws}\\
\mathcal{W}_{A}(\gamm)  & =\lim_{t\rightarrow0}~\frac{1}{2}\left(
\mathcal{K}_{u}(u)-\mathcal{K}_{s}(s)\right) =-\mathcal{K}_{s}(s)\,,\label{lwa}
\end{align}
the functions $\mathcal{W}_{S}(\gamm),\, \mathcal{W}_{S}(\gamm)$ are normalized
to unity in the nonrelativistic limit, i.e.,
$\mathcal{W}_{S}(1)=\mathcal{W}_{A}(1)=1$.

Vector and tensor integrals with four propagators are considered with respect
to the relations already obtained (\ref{int_b})-(\ref{s3}), (\ref{u3}). As an
example, consider an integral $(P_1=2p_1-q,\,P_2=2p_2+q)$:

\begin{eqnarray}
&& K^{\mathrm{(s)}}_\mu=\intk \frac{k_\mu}{(k^2-\la^2)
((k-q)^2-\la^2) (k^2-2p_1k)(k^2+2p_2k)} = \nonumber \\
&&= A P_{1\,\mu} + B P_{2\,\mu} + C q_\mu\,. \label{sv3}\\
\end{eqnarray}
Using the following relations for the logarithmic parts, we obtain the
coefficients $A$, $B$, $C$:

\begin{eqnarray}
&& q^{\mu} K^{\mathrm{(s)}}_\mu = \frac{q^2}{2}\,K^{\mathrm{(s)}}\\
&& P^{\mu}_1 K^{\mathrm{(s)}}_\mu = - J_2 - \frac{q^2}{2}\,K^{\mathrm{(s)}}\\
&& P^{\mu}_2 K^{\mathrm{(s)}}_\mu =   J_1 + \frac{q^2}{2}\,K^{\mathrm{(s)}},
\end{eqnarray}
here $J_1$ and $J_2$ are the integrals (\ref{int_tr}) with the propagators of
the first and the second particles, respectively;

\begin{eqnarray}
&& A =  \frac{4 s}{\Delta}\left[ \left(J_1 +
\frac{q^2}{2}\,K^{\mathrm{(s)}}\right) (P_1 P_2)+ \left(J_2 +
\frac{q^2}{2}\,K^{\mathrm{(s)}}\right)P_2^2\right]
\\
&& B = -\frac{4 s}{\Delta}\left[ \left(J_2 +
\frac{q^2}{2}\,K^{\mathrm{(s)}}\right)(P_1 P_2)+ \left(J_1 +
\frac{q^2}{2}\,K^{\mathrm{(s)}}\right) P_1^2
\right]\\
&& C = \frac{1}{2}\,K^{\mathrm{(s)}}\,,
\end{eqnarray}
where we introduce the following notation

\begin{equation}
\Delta = \frac{1}{4 s}\,\Big((P_1 P_2)^2-P_1^2 P_2^2\Big)\qquad
s=(p_1+p_2)^2\,. \label{delta}
\end{equation}

Tensor integrals can be considered in the same way.

\end{document}